\documentclass[aps,prx,superscriptaddress,twocolumn,amsmath,longbibliography]{revtex4-2}

\usepackage{graphicx}
\usepackage{dcolumn}
\usepackage{bm}

\renewcommand{\vec}[1]{{\boldsymbol #1}}
\usepackage[usenames,dvipsnames]{xcolor}

\begin{document}


\title{Topological insulator spin transistor}

\author{Linh T. Dang}
\affiliation{Physics Institute II, University of Cologne, D-50937 K{\"o}ln, Germany}

\author{Oliver Breunig}
\affiliation{Physics Institute II, University of Cologne, D-50937 K{\"o}ln, Germany}

\author{Zhiwei Wang}
\thanks{Present address: School of Physics, Beijing Institute of Technology, Beijing 100081, China}
\affiliation{Physics Institute II, University of Cologne, D-50937 K{\"o}ln, Germany}

\author{Henry F. Legg}
\affiliation{Department of Physics, University of Basel, CH-4056 Basel, Switzerland}

\author{Yoichi Ando} \email{ando@ph2.uni-koeln.de}
\affiliation{Physics Institute II, University of Cologne, D-50937 K{\"o}ln, Germany}


\begin{abstract}
When a charge current is injected into the surface state of a topological insulator (TI), the resulting shift of the spin-momentum-locked Fermi surface leads to the appearance of a net spin polarization. The helical spin structure of the Dirac-cone surface state of a TI should lead to a fixed sign of this spin polarization for a given current direction, but experimentally, both signs that agree and disagree with the theory expectation for the surface state Dirac cone have been observed in the past. Although the origin of the wrong sign has not been conclusively elucidated, this observation points to the possibility that one may switch the spin polarization at will to realize a spin transistor operation. Here we report the observation of both signs of spin polarization in the very same device and demonstrate the tunability between the two by electrostatic gating, which gives a proof of principle of a topological insulator spin transistor. This switching behaviour is explained using a minimal model of competing contributions from the topological surface state and trivial Rashba-split states. 
\end{abstract}

\maketitle

\newpage

\section{\label{sec:intro}Introduction}

A Datta-Das spin transistor \cite{Datta1990} can be an energy-efficient alternative of the traditional transistor and it represents one of the crucial components in spintronics \cite{RMP2004,nphys2007}. Instead of tuning the carrier density in a traditional field-effect transistor, in a spin transistor the gate acts on the spin degree of freedom of electrons. 
The main component of a Datta-Das spin transistor is a material with strong Rashba spin-orbit coupling \cite{Datta1990}. Significant efforts have been devoted to realize this type of spin transistor \cite{Chuang2015, PRL2021}, most notably on a gated 2-dimensional (2D) electron gas \cite{Chuang2015}. In this context, topological insulators (TIs) arise as a promising class of materials for spintronics \cite{Li2014, Ando2014, Tang2014, Liu2015, tian2015, dankert2015, Lee2015, Yang2016, Vaklinova2016, li2016, Li2019, He2022} due to the helical spin-momentum locking in their  surface state \cite{Ando2013} and spin transistor would be an interesting application of TIs \cite{Yang2016, Bandyopadhyay2022}.

When a charge current is injected into the surface state of a TI, the resulting shift of the Fermi surface leads to a net spin imbalance in the current-carrying electrons, which is called Edelstein effect \cite{edelstein1990}. This allows for generating a spin-polarized current without the need of a ferromagnetic (FM) spin source. In the past, multiple methods for the detection of this current-induced spin polarization (CISP) have been proposed and experimentally realized \cite{Li2014, Shiomi2014, yu2021,hus2017,leis2020}. Among these methods, using a FM potentiometer to probe the spin polarization as a spin voltage \cite{Johnson1988} is convenient, as it enables simple electrical detection of the gate-induced sign change of the CISP, which is the key signature of the spin transistor operation presented in this paper.

In the topological surface state (TSS) of TIs, the helicity of the spin polarization changes sign across the Dirac point (DP) \cite{Ando2013}. Hence, at first sight, one might expect that tuning the chemical potential across the DP by electrostatic gating can easily reverse the surface spin polarization in a current-biased TI. However, the Fermi velocity of the surface electrons also changes sign at the DP, which results in the same spin polarization for the same direction of the applied current. This mechanism has been explained in detail by Yang et al. \cite{Yang2016}. In other words, in TIs, the helicity of their TSS allows only one type of spin polarization that is determined by the sign of the applied current. Indeed, the expected sign of the CISP has been observed in many experiments \cite{Ando2014, Tang2014, tian2015, dankert2015, Lee2015, Yang2016, Vaklinova2016, li2016}. 

However, the opposite sign of the CISP has also been reported in TI devices \cite{Li2014,Yang2016,tian2021} and this must be explained by a different physical mechanism other than the current-carrying TSS electrons. The origin of this anomalous CISP has been a topic of interest and several possible explanations have been proposed, including bulk spin Hall effect \cite{tian2021}, Rashba-split states \cite{hong2012,Yang2016}, and spin-dependent interface resistance \cite{li2019a}. Unfortunately, no decisive evidence of any of these explanations have been found experimentally, leaving the origin of the opposite CISP ambiguous. Nevertheless, it has been remarked that devices with a higher carrier density (hence a higher chemical potential) tend to exhibit the unexpected CISP \cite{Yang2016, tian2021}.
Regardless of its origin, the existence of the CISP of opposite sign in devices with a higher chemical potential suggests that it may be possible to control the sign fo the CISP with electrostatic gating, thus realizing a spin transistor operation in a TI \cite{Yang2016}. 
So far, while a gate-tunability of the amplitude of the CISP (showing a peak near the DP) has been reported \cite{Lee2015}, the gate-induced switching of the CISP has not been achieved \cite{Lee2015,tian2021}.   

\begin{figure*}[tb]
\includegraphics[width=0.75\linewidth]{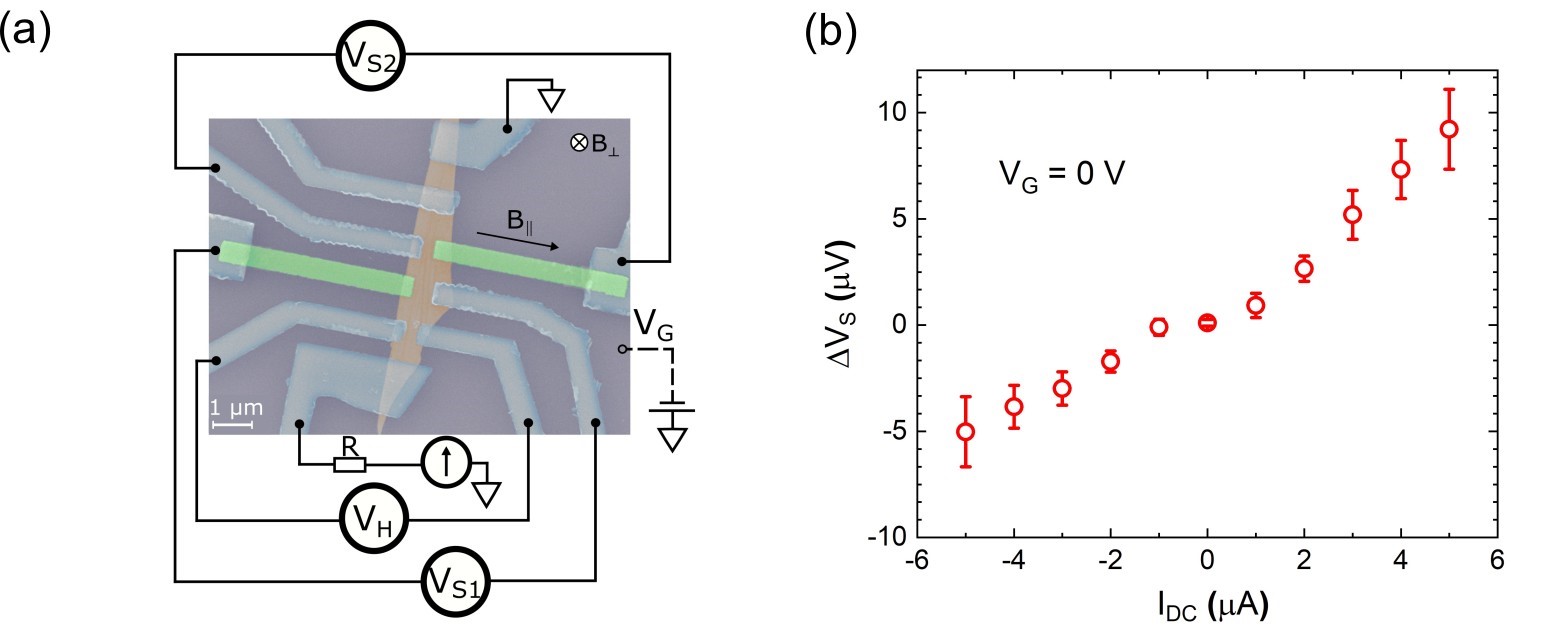}
\caption{\linespread{1.05}\selectfont{}
\label{fig:1}Topological insulator spin transistor device. a) False-colored SEM image of device A with the measurement scheme. Normal contacts are fabricated using Nb (shaded in light blue). The tunnel barrier between the Permalloy (Py) leads (light green) and the BSTS flake (light orange) is formed by a layer of SiN$_x$ deposited right after exfoliation of the flake. A direct current $I_\mathrm{DC}$ is injected from the bottom Nb lead via a 1-M$\Omega$ resistor and drained at the top Nb lead. The voltages between the neighbouring Nb and Py leads, $V_\mathrm{S1}$ and $V_\mathrm{S2}$, are acquired while sweeping the magnetic field $B_{\parallel}$ aligned along the long axis of the Py leads. An out-of-plane magnetic field $B_{\perp}$ was applied for Hall measurements. The gate voltage $V_\mathrm{G}$ is applied to the back side of the degenerately doped Si substrate which is covered by a 290 nm of SiO$_2$ acting as a gate dielectric. b) The spin voltage $\Delta V_\mathrm{S}$ (open symbols) shows an approximate linear dependence on the applied current $I_\mathrm{DC}$ and its sign inverts upon reversal of the current direction. The data shown were measured on device A at $V_\mathrm{G}$ = 0 V.}
\end{figure*}

For the purpose of controlling the sign of the spin voltage with electrostatic gating, in this study we base our devices on TI flakes that were exfoliated to the thicknesses of less than $\sim$20 nm in order to achieve a sizeable tunability of the top TSS from bottom gating. Secondly, we used bulk crystals of Bi$_{2-x}$Sb$_{x}$Te$_{3-y}$Se$_{y}$ (BSTS) with different compositions to obtain different chemical potentials at zero gate voltage. Upon gating, we observed a maximum in the CISP near the DP in one device and sign switching of the CISP, with complicated multiple-switching behaviour, in two other devices. Furthermore, we provide a theoretical explanation of this switching behaviour based on the assumption that the opposite sign of the CISP originates from trivial Rashba-split states on the surface of a TI.

\section{\label{sec:RAD}Experimental Results}

\subsection{Devices for spin voltage detection}

For all of the analyzed devices, the voltage $V_\mathrm{S}$ measured between the Py lead and its reference lead, which reflects the relative orientations of the spins in Py and in the TI surface, shows a clean hysteresis as a function of $B_{\parallel}$ with a coercive field of about 20 Oe as expected for a thin ($\sim$30 nm) Py strip that is magnetized along its easy axis (see Fig. 1a for device schematics). In contrast to previous experiments where averaging of multiple sweeps was required to extract the hysteresis \cite{Ando2014, Yang2016}, here, we could resolve it even from a single magnetic field sweep due to an enhanced signal-to-noise ratio and reduced charge fluctuations. The latter owes to a better tunnel barrier made of SiN$_x$ (see Methods section for details). 
The size of the voltage hysteresis, i.e. the spin voltage $\Delta V_S$ which is proportional to the CISP on the TI surface, shows an approximate linear dependence on the applied current (Fig. 1b) and changes sign upon reversal of the current direction, as expected. The spin resistance $R_\mathrm{S} = \Delta V_\mathrm{S}/I = 1.18$ $\Omega$ in device A is similar to previously reported values in the range of 1--5 $\Omega$ \cite{li2016, li2019a, tian2015}

\begin{figure*}[t]
\includegraphics[width=\linewidth]{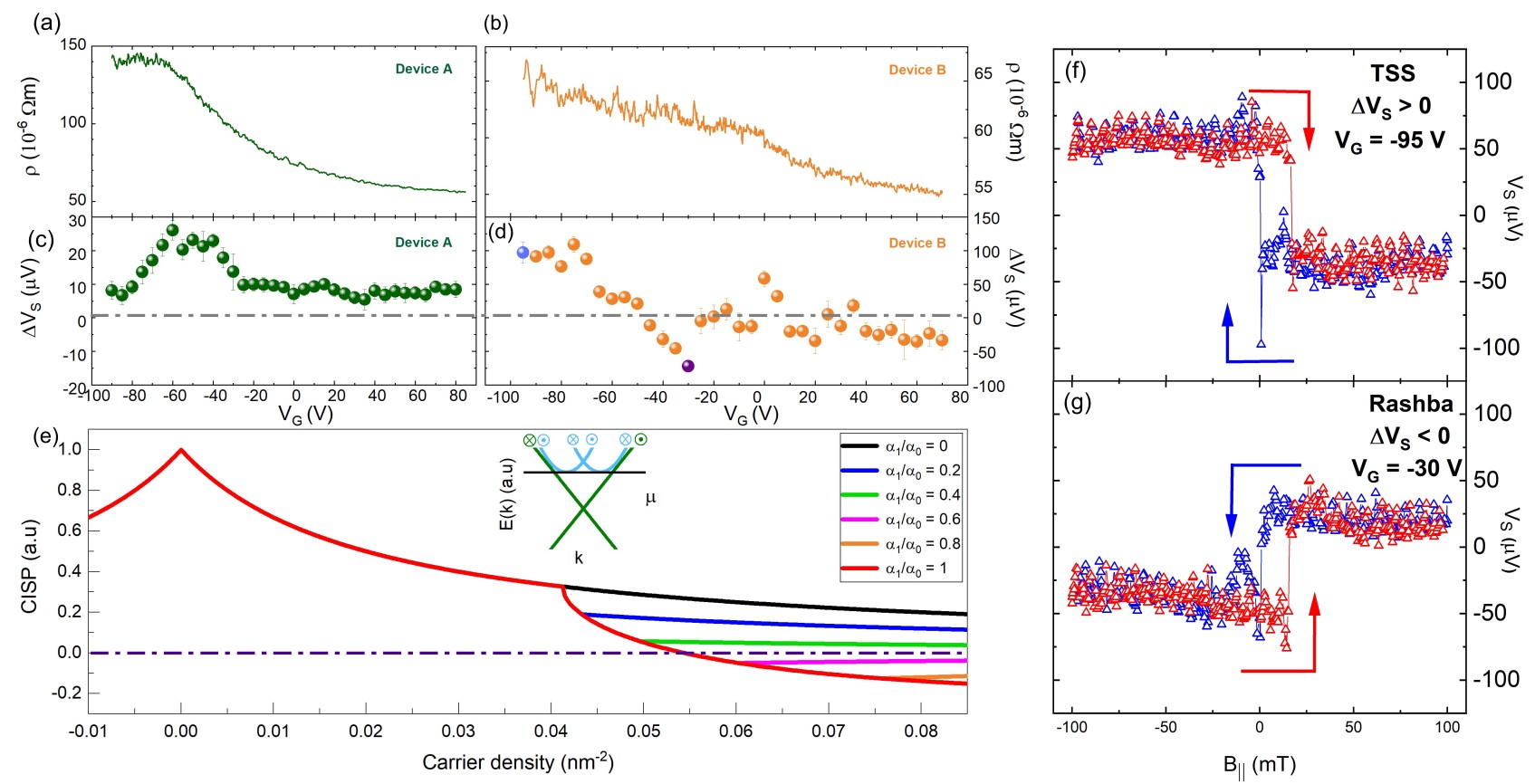}
\caption{\linespread{1.05}\selectfont{}
\label{fig:2} Gate-controlled spin voltage. a), b) Resistivity $\rho$ as a function of gate voltage $V_\mathrm{G}$ for device A and B, respectively, measured in a 4-terminal configuration.  c), d) Spin voltage $\Delta V_\mathrm{S}$ as a function of $V_\mathrm{G}$ in device A and B, respectively. 
In device B, multiple sign changes are observed. The dashed-dotted gray line denotes zero spin voltage for both devices. 
e) Theoretically calculated carrier-density dependence of the current-induced spin polarization (CISP) for different ratios of the contributions from the Rashba band and TSS, $\alpha_{1}/\alpha_{0}$. The CISP has a global peak at the DP and an abrupt change when the chemical potential reaches the Rashba band. With  $\alpha_{1}/\alpha_{0} > 0.4$, the total CISP can turn negative (zero marked by the dashed-dotted line). The parameters used are $\alpha_0=250$ meV~nm (which is consistent with $v_F\approx 4\times10^5$ m/s of the TSS), $m_1=300$ meV~nm$^2$ (which gives a $v_F$ similar to that of the TSS) and $\mu_1=180$ meV (close to the bulk conduction band edge); the behaviour of the CISP is, however, only weakly dependent on the exact choice of parameters.
Inset: Schematics of a TSS (green) with a Rashba band (blue) at about 200 meV away from the DP. The momentum-locked spin of electrons in each branch is denoted on top. The chemical potential (black line) crossing the Rashba band bottom corresponds to the kink in the CISP at a carrier density of about 0.04 nm$^{-2}$. f), g) Exemplary raw data of the hysteresis observed in the $V_{\rm S}$ vs $ B_{\parallel}$ behaviour at $V_\mathrm{G}$ of $-95$ and $-30$ V; the corresponding $\Delta V_\mathrm{s}$ values extracted from these data are shown in panel d) with blue and purple symbols, respectively.}
\end{figure*}

Next, we turn to the dependence of the spin voltage on the chemical potential (gate voltage) to study the different signs that have been reported. Here, for the first time we observed both signs of the spin voltage in a single device upon changing the gate voltage (Figs. 2f and 2g). However, before discussing the devices presenting the sign reversal, we first consider the reference device A (Fig. 2a). The TI flake used for device A was bulk-insulating and the chemical potential can be tuned to the DP with a gate voltage $V_{\mathrm{G},\mathrm{peak}}^\rho \approx -75$ V, as reflected in a resistance peak of the longitudinal resistivity $\rho$. Similarly, $\Delta V_\mathrm{S}$ also presents a maximum, but at a slightly different gate voltage $V_{\mathrm{G}, \mathrm{peak}}^\mathrm{S} \approx -60$ V (Fig. 2c). 
Even though $\Delta V_\mathrm{S}$ is expected to be maximum when the chemical potential is at the DP \cite{Lee2015}, a finite difference between $V_{\mathrm{G},\mathrm{peak}}^\rho$ and $V_{\mathrm{G},\mathrm{peak}}^\mathrm{S}$ may originate from the fact that both the top and bottom TSSs contribute to the resistance of the flake, while only the top TSS contributes to the spin voltage. These two TSSs experience different chemical potential tuning by the gate. 
Tuning the gate away from the spin voltage peak, $\Delta V_\mathrm{S}$ decays quickly and saturates for $V_\mathrm{G} \gtrsim -30$ V, which fits well with the expectation for a single Dirac cone contribution to spin polarization \cite{li2016interp,leis2020}. 
From an estimation of the geometric capacitance of the gate for this device, the conduction band edge is expected to be reached at a gate voltage $V_\mathrm{CB}\approx 60$ V. Since the Rashba-split band is formed very close to the conduction band edge, one expects the onset of its contribution at this voltage. The absence of clear features in $\Delta V_\mathrm{S}$ around 60~V is explained by screening effects or by an electric field-induced suppression of the Rashba splitting due to the large gate voltage that is required to tune to the conduction band (see below for details).

\subsection{Gate-controlled sign reversal of the spin voltage}

To better access the Rashba-split band, another device B was fabricated from a more n-type TI flake, such that a smaller gate voltage suffices to tune the chemical potential to the conduction band. In this device, the spin voltage indeed shows a much richer gate response (Fig. 2d). The $V_{\rm G}$ dependence of $\rho$ in this device (Fig. 2b) shows n-type behaviour with smaller absolute values of $\rho$ and a reduced gate tunability, both indicating that the chemical potential in this device without gating is closer to the conduction band. Comparing the resistivity values of this device to device A, a rough estimate of a gate voltage shift between device A and device B of about 50 V can be deduced.
Consistent with this estimate, the resistance cannot be tuned to a peak within the gate voltage range accessible in our experiment and the device remains n-type.
Interestingly, the spin voltage $\Delta V_\mathrm{S}$ of device B shows multiple sign switches (below we  show we can attribute a positive or negative sign of $\Delta V_\mathrm{S}$ to the TSS or trivial Rashba bands, respectively). At large negative gate voltages, the positive sign is consistent with a dominant TSS contribution. Upon sweeping $V_{\rm G}$ to the positive direction, however, $\Delta V_\mathrm{S}$ drops abruptly near $-70$ V and a sign reversal occurs close to $-50$ V, indicating the appearance of another contribution and eventual reversal of the spin polarization. With further sweeping of $V_{\rm G}$ to the positive direction, the sign of $\Delta V_\mathrm{S}$ switches a couple of times and eventually converges to a negative value at $V_\mathrm{G} \gtrsim$ 40~V. Even though $\Delta V_\mathrm{S}$ seemingly fluctuates between plus and minus, it should be noted that for a given fixed gate voltage, the hysteresis loops such as shown in Figs. 2f and 2g are static and reproducible. Yet, their sign can invert quickly with a small change in the gate voltage. In other words, the $V_\mathrm{G}$ dependence of $\Delta V_\mathrm{S}$ reflects an intrinsic property that changes rapidly within a certain gate-voltage range and it is not the result of a random fluctuation or instability. Here, the opposite spin polarization was observed at different gate voltages in the same device, which has not been reported before in this type of TI-based devices. 

Notably, the spin voltage reversal was observed in the more n-type device B, while it was completely absent in device A. The absence of Rashba contributions upon gating in device A could be explained in two ways. First, due to the screening effect from the FM lead, the chemical potential underneath the lead is not efficiently tuned by the electrostatic gate. Therefore, even at a gate voltage high enough to bring a crystal of similar pristine carrier density to the conduction band, the chemical potential in device A may still resides in the bulk gap. Second, it is known that a strong electric field can affect the splitting of a Rashba-split state \cite{caviglia2010}, effectively changing (or even reversing) its contribution to the spin polarization. In device A, the chemical potential is closer to the DP without gating, hence a larger gate voltage (i.e. a large electric field) is required to bring the chemical potential to the conduction band edge. This large electric field can then weaken the contribution from the Rashba band to the spin polarization, leading to the absence of an opposite spin polarization in device A. Both mechanisms can be circumvented by employing a crystal with a higher pristine chemical potential. Indeed, device B as well as device C (see Supplemental information), which were fabricated from such a crystal (see Methods), exhibit the Rashba-type contribution, i.e., the sign reversal of $\Delta V_\mathrm{S}$. 

\subsection{Effects of chemical potential pinning by Py and spin diffusion}

One of the previous explanations of the unexpected sign of the spin voltage was based on the contribution of the trivial Rashba-split states formed underneath the metallic ferromagnetic contacts due to band bending \cite{Yang2016}. However, in this scenario one would expect that the chemical potential underneath the metal contacts, i.e. the spin detector, is fixed and cannot be changed with the gate. Thus, at first sight, one might not expect any gate dependence of the spin voltage from this area at all. However, our data clearly shows that the spin voltage detected is gate-tunable: Device A exhibits a peak in $\Delta V_\mathrm{S}$ near the DP, indicating that the chemical potential of the spin-polarized electrons is indeed modified and can be brought to the DP. If the CISP would originate from electrons underneath the FM leads alone, this should not happen. This contradiction may be resolved by considering spin diffusion from the (free) area not covered by the FM leads to the area underneath the FM leads. These two areas differ in terms of the electronic states and the gate tunability:
In the free area, the chemical potential is not pinned, but can be tuned by the bottom gate via the electrostatic coupling between the top and bottom TSSs \cite{fatemi2014}. In addition, Rashba-split states due to trivial surface states can be formed in the free area \cite{bianchi2010,zhu2011,king2011,bahramy2012}. Those may become visible in the CISP probed by the FM detector via spin diffusion. 
In the area beneath the FM lead, Rashba-split states would also exist due to band bending effects at the Py/TI interface and, even though the chemical potential is pinned here, the size of the Rashba splitting is modified by the electric field generated upon gating, leading to a modification of the CISP in this area as well \cite{caviglia2010, Yang2016}. 

The spin voltage detected in the experiment reflects all of these sources of spin polarization via averaging the CISP in an area that is larger than the detector by an area set by the order of the spin diffusion length.
Typical spin diffusion lengths in TI materials (up to 1~$\mu$m, \cite{iyer2018}) are of similar order as the size of the present FM detector, such that the observed CISP can be understood as simply the average of the contributions from the two areas.

\begin{figure}[tb]
\includegraphics[width=0.9\linewidth]{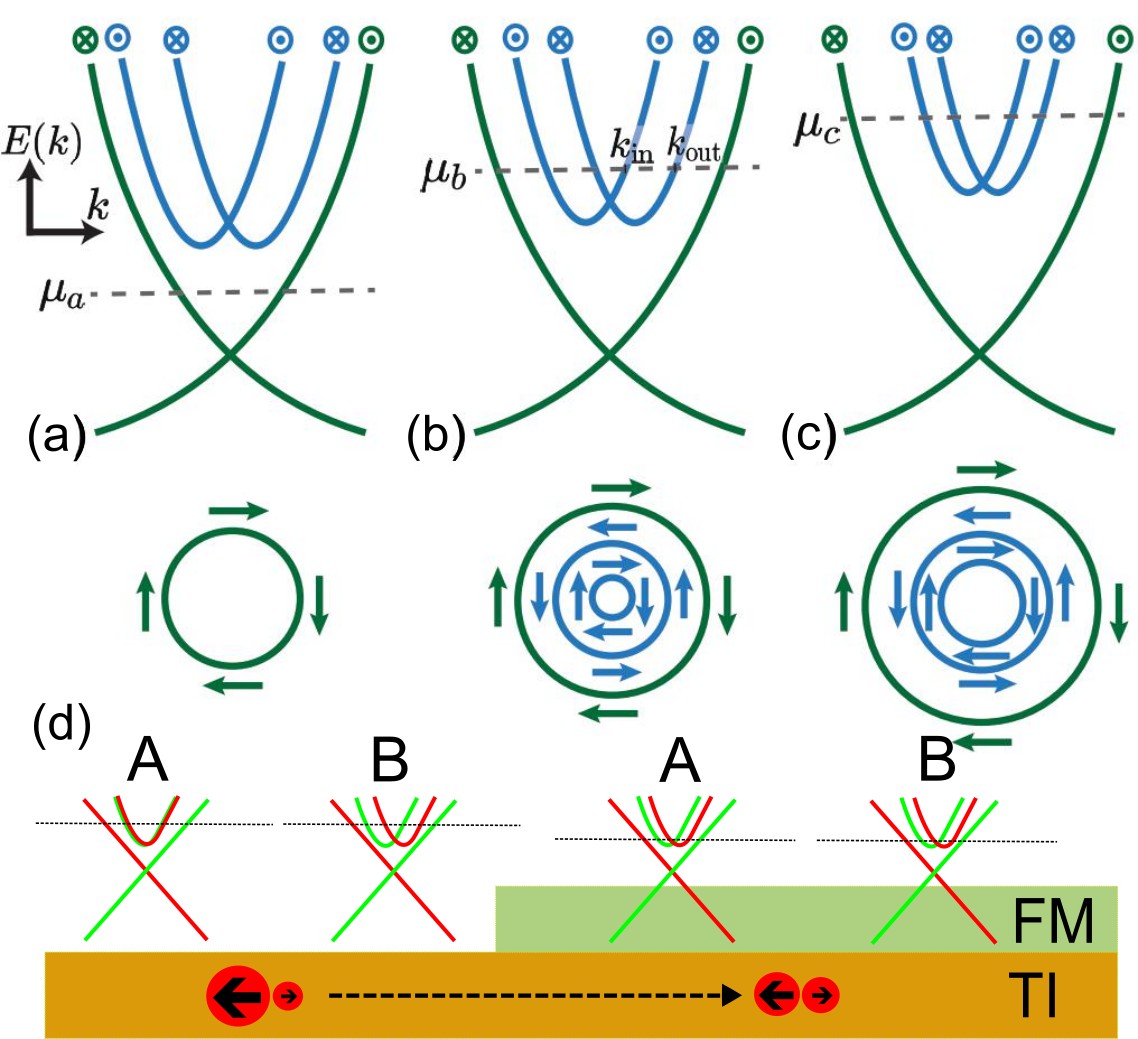}
\caption{\linespread{1.05}\selectfont{}
\label{fig:3} Schematics of the surface bands during gating. a) When the chemical potential is located close to the Dirac point, the CISP arises solely from the TSS. b) As the chemical potential is raised close to the bulk conduction band, trivial Rashba surface bands contribute to the CISP; here, only a single band is shown. The contribution of trivial bands has opposite sign to the TSS  contribution and can cause a sign change in the total CISP. c) As the chemical potential is tuned to a higher value, the contribution of the Rashba band can be further modified; this could occur, for example, due to the electric field generated by gating (shown) or the momentum dependence of the Rashba coefficient $\alpha_i$ (not shown). As a result, the CISP can change sign multiple times depending on the relative values of the trivial and topological CISP contributions. d) Two parts of the device: free TI surface, left, and the surface covered by the FM lead (green), right. The band structures of these two parts are sketched on top of them. Both of them have spin polarized (red and green) bands: TSS and Rashba-split states. The chemical potentials and the Rashba splittings are different between the two areas, resulting in different CISPs (red circles with arrows denote the spin direction). Electrons with different spin polarization then diffuse from one part to another. The sketch is for a finite gate voltage applied, which has tuned the chemical potential into the Rashba bands on the left, while it is pinned on the right side. Labels A and B refer to the scenarios in devices A and B. Note the different size of the Rashba splitting between devices A and B in the free area.}
\end{figure}

\section{Theoretical understanding of the spin transistor effect}

The mechanism of the spin transistor effect is graphically demonstrated in Fig. 3d, which depicts the different behaviour of devices A and B. In device A, for bringing the chemical potential into the Rashba states of the free area, a very large gate voltage is required, which significantly reduces the Rashba coefficient such that there is no sizeable contribution of Rashba-type to the detected CISP. Additionally, the Rashba states formed underneath the FM may contribute as well, but the magnitude is mostly set by the details of the FM/TI interface and the tunability by the gate is limited.
In the more n-type device B (and in device C, see Supplementary Information) the Rashba splitting in the free area remains large enough upon gate tuning to induce a negative contribution to the CISP.


The key features of the CISP observed in our experiment can be described with a minimal model that includes the contributions from both the topological surface state and surface Rashba states. We consider a simple Hamiltonian for the surface of our TI, such that
\begin{align}
H=\sum_i \tau_i \left[\frac{\vec k^2}{2m_i}+\alpha_i (\sigma_y k_x - \sigma_x k_y)-\varepsilon_i-\mu\right],
\end{align}
here $\sigma_x$ and $\sigma_y$ are Pauli matrices in spin-space, $\tau_i$ is a diagonal matrix in band-space that distinguishes the $i$th band, $m_i$ is the effective mass of this band, $\varepsilon_i$ governs the $k=0$ crossing point of this band, and $\mu$ is the chemical potential. In particular, we label the topological surface state by $i=0$ such that $m_0^{-1} =0$ and measure the chemical potential from the Dirac point such that $\varepsilon_0=0$. Note that throughout we set $\hbar=1$. Importantly, we set the $\alpha_i$ parameters of this model all of the same sign, such that the Rashba parameters of the trivial surface states are consistent with the sign of the helicity of the topological surface state (see Fig. 3).

Within this simple model, the CISP due to an electric field $\mathcal{E}_x$ can be written as \cite{li2016} (see Methods for full details)
\begin{equation}
P_{\rm CI}=\frac{n_\uparrow-n_\downarrow}{n}=e \mathcal{E}_x\frac{\sum_{i} \Delta_i }{4\pi n},
\end{equation}
where $n_{\uparrow(\downarrow)}$ is the spin-up(down) density and $n=n_{\uparrow}+n_{\downarrow}$ is the total density. The contribution from the topological surface state $(i=0)$ is 
\begin{equation}\label{topeq}
\Delta_{0} = \tau_0 k_{F,0}  \approx \frac{8 \pi  \alpha_0}{\gamma},
\end{equation}
where $\gamma$ is the impurity strength. Similarly, the Rashba surface state ($i\geq1$) contributions are of opposite sign and take the form
\begin{equation}
\Delta_i = - \tau_i \left(k_{F,i}^{\rm out}-k_{F,i}^{\rm in} \right) \approx  -\frac{8 \pi}{m_i \gamma} \left(k_{F,i}^{\rm out}-k_{F,i}^{\rm in} \right),
\end{equation}
where $k_{F,i}^{\rm out}$ ($k_{F,i}^{\rm in})$ are the outer (inner) Fermi-momenta of the $i$th Rashba surface band (see Fig. 3). We also defined the scattering time for each band, $\tau_i$, and in the approximation we assume that only intraband scattering is relevant, such that $\tau_i$ is governed by the impurity strength, $\gamma$, and the density of states of the band $i$. 

The situations of the bands reflecting the effects of increasing gate-induced electric field are depicted in Figs. 3a-3c. As expected, when the chemical potential is in the vicinity of the Dirac point, the CISP is entirely dominated by the contribution of the surface state. However, when the chemical potential is tuned close to the edge of the bulk conduction band, trivial Rashba surface state(s) result in a contribution to the CISP which has the opposite sign and therefore can cause a total sign change of the CISP if $\Delta_1>\Delta_0$ (see Fig. 2e).

Finally, we comment on how the full experimental behaviour, including multiple sign changes far from the Dirac point, can be understood within this simple model: If the contribution of the Rashba states, $\Delta_{i>0}$, are dependent on the chemical potential, several sign changes can occur in the vicinity of the bulk conduction band as the gate voltage is varied (see Fig.~\ref{fig:3}). In fact, such a dependence is quite natural. For instance, the Rashba strength $\alpha_i$ can be modified by the electric field that arises due to gating \cite{caviglia2010} and/or the coefficient $\alpha_i$ will in reality be dependent on momentum $k$. Furthermore, the mass $m_i$ will have momentum dependent corrections that could also reduce the magnitude of $\Delta_i$. In contrast, it is known that the Fermi velocity, $\alpha_0$, of the topological surface Dirac cone increases close to the bulk conduction band 
and this will lead to a corresponding increase in the contribution $\Delta_0$, see Eq.~\eqref{topeq}. Hence, the relative magnitudes of the topological and trivial CISP contributions can explain the multiple switches observed experimentally. The full understanding of this multiple switchings behaviour is an important subject for the future applications of topological insulator spin transistors.

\section{Conclusion}

We studied the gate-voltage dependence of CISP in different devices fabricated from Bi$_{2-x}$Sb$_{x}$Te$_{3-y}$Se$_{y}$ crystals. We observed gate-controlled sign reversals of the CISP for the first time in TI devices and such sign reversals were found to occur only in devices fabricated from n-type crystals. 
We argue that trivial Rashba-split states on the TI surface contributes to the CISP with an opposite sign compared to that from the TSS and the dependence of the Rashba parameter $\alpha$ on the electric field generated by the back gate plays a crucial role in their competition, leading to complicated sign reversals in the observed CISP as a function of the gate voltage.  Our devices demonstrate the feasibility of the spin transistor conceived by Yang {\it et al.} in Ref. \cite{Yang2016}, while presenting the importance of spin diffusion in its working principle. The spin transistor effect demonstrated here will significantly widen the prospect of TIs for spintronic applications. For fundamental research, the miniaturization of the spin transistor into the ballistic transport regime would be extremely interesting, because the efficiency of the CISP would be drastically enhanced compared to the diffusive transport regime \cite{Ando2013}.

\section{Acknowledgements}
This work has received funding from the Deutsche Forschungsgemeinschaft (DFG, German Research Foundation) under CRC 1238-277146847 (subprojects A04 and B01) and under 398945897, as well as under Germany’s Excellence Strategy -- Cluster of Excellence Matter and Light for Quantum Computing (ML4Q) EXC 2004/1-390534769.
H. F. Legg acknowledges support by the Georg H. Endress Foundation.

\vspace{3mm}

\begin{flushleft}
{\bf Apendix}
\end{flushleft}

\begin{flushleft}
{\bf Device fabrication:}\\
\end{flushleft}
\vspace{-3mm}
The data shown in this paper were taken on three different devices A,B and C. Two different TI crystals were used for device fabrication: device A was fabricated from BiSbTeSe$_{2}$ bulk crystal, while device B and C were fabricated from Bi$_{1.5}$Sb$_{0.5}$Te$_{1.3}$Se$_{1.7}$ crystal with a chemical potential closer to conduction band edge compared to the former. Both single crystals were grown by the modified Bridgman method.
TI flakes were exfoliated from single crystals using standard scotch tape technique onto degenerately doped Si substrates coated by a 290-nm layer of SiO$_{2}$ acting as a gate dielectric. Thin flakes ($<$ 20 nm) with a suitable shapes and sizes and flat surfaces were selected using optical microscopy. 
After exfoliation the substrate was immediately loaded to the vacuum chamber of the HW-CVD machine to minimize any native oxide formation on the TI surface. There, a 1-nm-thick layer of SiN$_x$ was deposited on top of TI, acting as a capping layer and tunnel barrier at the same time. During the SiN$_x$ deposition, the substrate temperature was observed to stay below 100$^{\circ}$C. Diluted ZEP 520A resist was spin coated immediately after the SiN$_x$ deposition. Electron beam lithography was performed using a Raith Pioneer Two system. Contacts were fabricated using Niobium (Nb) and Permalloy (Py) for the normal and FM contacts, respectively, by a liftoff-technique. 
For normal leads, first the SiN$_x$ layer was removed using reactive ion etching in CF$_{4}$ plasma, followed by the deposition of 35-nm-thick Nb via magnetron sputtering. In a second lithography, 30-nm-thick Py leads were deposited by thermal evaporation without removing the SiN$_x$ layer.
The tunnel-barrier resistances of the measured FM leads of device A, B and C were 4.2, 4.1 and 6.4 $\text{k}\Omega$, respectively.  
Finally, 40-nm-thick Nb leads were fabricated to connect to pre-patterned Au bonding pads.

\begin{flushleft}
{\bf Measurements:}\\
\end{flushleft}
\vspace{-3mm}
All measurements were performed in Oxford Instruments TRITON 200 dilution refrigerator at the base temperature of 20--40 mK. A vector magnet was used to allow measurement of multiple devices with different orientations in the same run. 
The wiring scheme of the spin voltage measurement is illustrated in Fig. 1a. DC current is applied using the two outer Nb contacts and an in-plane magnetic field $\lvert B_{\parallel}\rvert\leq 100$ mT is applied along the easy axis of the FM contact to switch its magnetization. The voltage $V_{S}$ is measured between a FM lead and the opposing Nb lead as a function of the magnetic field using a Keithley 2182A nanovoltmeter; this contact arrangement minimizes the contribution of the longitudinal resistance. Further data treatments to subtract the remaining longitudinal resistance contribution is described in the Supplementary Information. The back-gate voltage is applied to the sample using a Keithley 2450 sourcemeter.  Each measurement includes multiple forward (negative $B_{\parallel}$ to positive $B_{\parallel}$) and backward (positive $B_{\parallel}$ to negative $B_{\parallel}$) magnetic-field scans. An average $\Delta V_{S}$ is deduced from multiple scans for each biased current or gate voltage to reduce the noise.  
The longitudinal resistance was measured in a standard 4-probe configuration with a low-frequency (13.77 Hz) AC technique using a NF LI5640 lock-in amplifier.

\begin{flushleft}
{\bf Calculation of the current-induced spin polarization:}
\end{flushleft}
\vspace{-2mm}
First, to connect with the experimental gate voltage dependence of the CISP, we make the simplifying assumption that the capacitance is independent of chemical potential and that the gate voltage changes only the surface density. Using Luttinger's theorem we obtain this density, such that
\begin{equation}
n=\frac{k_{F,0}^2}{4\pi}+\sum_{i>0} \frac{(k_{F,i}^{\rm out})^2+(k_{F,i}^{\rm in})^2{\rm sgn}(\mu-\varepsilon_i)}{4\pi}+n_{\rm bulk},
\end{equation}  
where $n_{\rm bulk}$ encodes the fact that in any real experimental setup there is a contribution from bulk impurity states to the density $n$, this prevents the divergence of the CISP when the chemical potential is at the Dirac point. This equation can be used to obtain $\mu(n)$ because we calculate the CISP (below) for a given chemical potential.

We now derive Eq.~(2), namely, the contributions of each individual Fermi-pocket (shown in Fig. 3b) to the CISP. Using the Boltzmann description \cite{li2016}, 
we can write the occupation function in the presence of an electric field $\mathcal{E}_x$, to linear order, as $f^{(i)}(\vec k)\approx f^{(i)}_0(\vec k)+e \tau_i \mathcal{E}_x \frac{\partial f^{(i)}_0(\vec k)}{\partial k_x}$. For a given chemical potential the contribution to the spin density of a Fermi surface with helicity $\xi=\pm1$ is given by
\begin{equation}
n_{\uparrow (\downarrow),i}=\sum_{j\in\{{\rm in},{\rm out}\}}\frac{1}{8\pi}\left((k^{j}_{F,i})^2+ (-) \xi_j k^j_{F,i} e \tau_i \mathcal{E}_x\right),
\end{equation}
where the inner and outer Fermi-pockets always have opposite helicity (see Fig.~3).
Hence the difference in spin density for a single pair of bands $i$, as appears in the CISP, is given by
\begin{equation}
n_{\uparrow,i}-n_{\downarrow,i} = -e  \mathcal{E}_x \frac{ \tau_i (k^{\rm out}_{F,i}-k^{\rm in}_{F,i})}{4\pi}= \frac{e \mathcal{E}_x}{4\pi} \Delta_i,
\end{equation}
where for the topological band $k^{\rm out}_{F,i}=0$ because there is only the inner Fermi-pocket.

Finally, to calculate the CISP (the result of which is shown in Fig. 2e), we need to approximate the transport scattering time, $\tau_i$, for the $i$th pair of bands. We do this by assuming only intra-Fermi pocket scattering is relevant, such that the scattering rate is given by
\begin{equation}
\frac{1}{\tau_{i}}= \gamma  \nu_i(\mu) \int \frac{d\theta}{2\pi} (1-\cos \theta) \cos^2 \frac{\theta}{2}=\frac{\gamma \nu(\mu)}{4},
\end{equation}
where $\nu_i(\mu)$ is the density of states of the given Fermi pocket for the $i$th band. For the topological Fermi pocket $\nu_0(\mu)=\frac{\mu}{2\pi\alpha_0^2}$ and for trivial Fermi pocket $\nu_{i\neq0}(\mu)=\frac{m_i}{2\pi}$. Inserting these scattering times into $\Delta_i$ gives the approximations in Eq.~(3) and Eq.~(4). In reality, inter-pocket scattering will also occur and this would couple all bands. However, our simple approximation using only intra-pocket scattering is sufficient to capture the change of sign in the CISP observed in our experiment.

\providecommand{\noopsort}[1]{}\providecommand{\singleletter}[1]{#1}%

\end{document}


\maketitle

\section{Device summary}
A total of 9 batches with a SiN$_x$ tunnel barrier were fabricated (each batch was a set of tens of devices fabricated on the same Si/SiO$_2$ chip, going through the same fabrication process). Different batches of fabrications were performed to optimize the SiN$_x$ tunnel barrier for a suitable tunnel resistance, which we obtained in 2 of these batches. From the batches with optimized SiN$_x$ fabrication, we measured 12 devices and the spin voltage was observed in 3 of them: device A came from a batch using BiSbTeSe$_{2}$ crystal while device B and C came from another batch using Bi$_{1.5}$Sb$_{0.5}$Te$_{1.3}$Se$_{1.7}$ crystal.    

\section{Estimation of spin voltage}
\begin{figure}[h]
    \centering
    \includegraphics[width=0.9\linewidth]{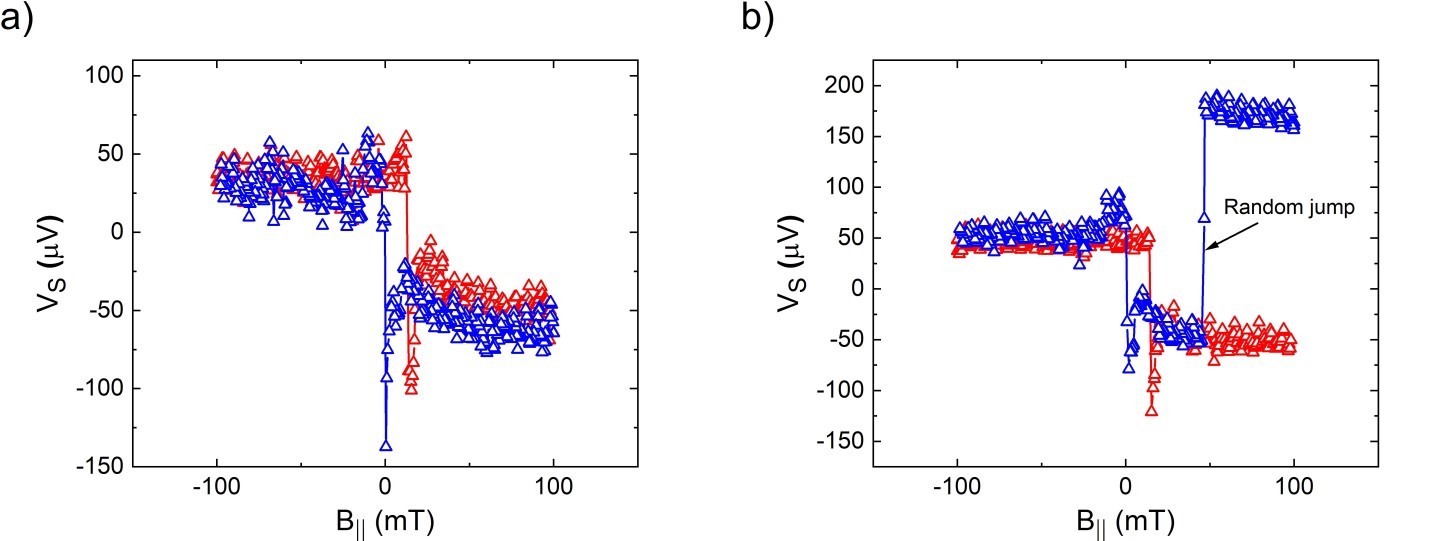}
    \caption[width=\linewidth]{\linespread{1.05}\selectfont{}\label{fig:S1} a) Scan without a random jump in the signal. b) A random jump in the signal observed during the magnetic-field scans in the same measurement setting. Multiple scans are performed for each $V_{\rm G}$. }
\end{figure}

For each data point in the plots of $\Delta V_\mathrm{S}$ vs $V_\mathrm{G}$, the bottom gate voltage $V_\mathrm{G}$ is first set at the targeted value, then a bias current of 5 $\mu$A is applied through the device, followed by a magnetic field sweep applied perpendicular to the current direction. The magnetic field is first ramped up to $B_\mathrm{max}$, then swept between $\pm B_\mathrm{max}$ five times before ramped back to 0. Each of these five down-up scans is then analyzed to extract $\Delta V_\mathrm{S}$. However, sometimes a sudden, irreproducible jump in the signal occurred during the scan, whose origin is unclear; such a random jump typically happened in 1 or 2 scans among the 5 scans recorded for each $V_\mathrm{G}$, such that there were usually a sufficient number of scans at each $V_\mathrm{G}$ to extract $\Delta V_\mathrm{S}$ without being affected by a random jump in the signal.  Figure \ref{fig:S1} demonstrates two scans, with and without a random jump. 
Among the five scans at each $V_\mathrm{G}$, only those scans in which the random jumps in the signal do not hinder the determination of a stable high-field value of $V_\mathrm{S}$ were analyzed.

Prior to the estimation of $\Delta V_\mathrm{S}$, a linear background subtraction process is performed on each gate-voltage scan set. Due to fabrication mismatch between the Py lead and its Nb counterpart, $V_\mathrm{S}$ always contains a small longitudinal Ohmic voltage drop. This Ohmic contribution can have a magnetic field dependence (due to the magnetoresistance of TI) and create spurious features in the measured voltage. In order to remove this contribution, a longitudinal voltage $V_\mathrm{L}$ was measured between two Nb leads (for example the two outer Nb leads in Fig. 1 of the main text) during the magnetic field scan. This $V_\mathrm{L}$ was then normalized by the zero-field value: $V_\mathrm{L(norm)}=V_\mathrm{L}/V_\mathrm{L(B_{\parallel}=0)}$. This normalized change of the longitudinal voltage was subtracted from $V_\mathrm{S}$ to obtain the background-subtracted $V'_\mathrm{S}$ as follows:
\begin{equation}
    V'_\mathrm{S} = V_\mathrm{S} - V_\mathrm{L(norm)}\times V_\mathrm{S(B_{\parallel}=0)}.
\end{equation}

\begin{figure}[h]
    \centering
    \includegraphics[width=0.5\linewidth]{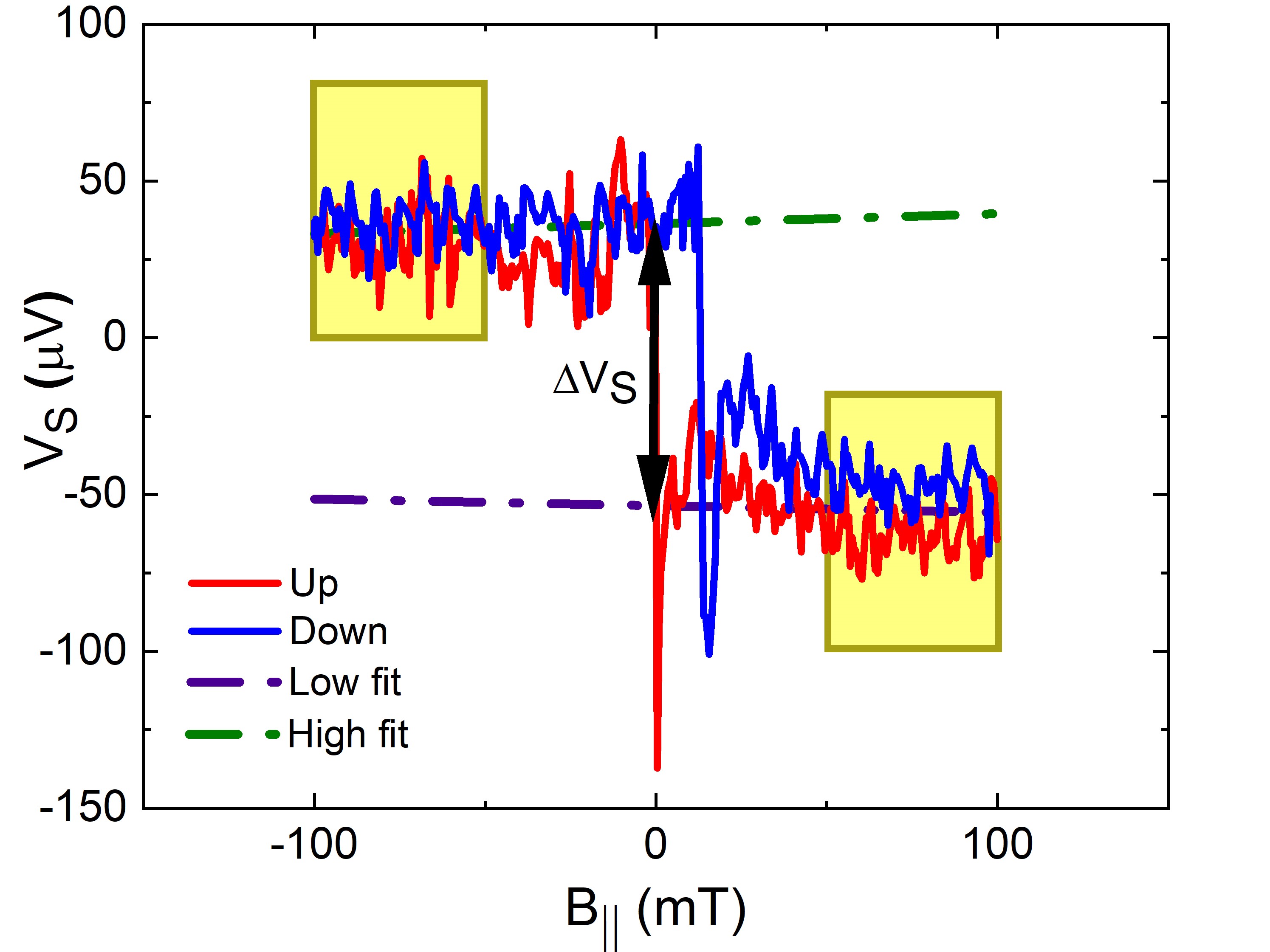}
    \caption[width=\linewidth]{\linespread{1.05}\selectfont{}\label{fig:S2} Analysis of the data to extract the spin voltage.  Light yellow box denotes the data range for fitting. Green and purple dash-dotted lines are the fitted lines for high/low potential, respectively. The spin voltage $\Delta V_\mathrm{S}$ is the difference in the intersections of these lines on the $V_\mathrm{S}$ axis at $B_{\parallel}$ = 0 T.}
\end{figure}

Afterwards, in order to extract $\Delta V_\mathrm{S}$ from a single scan, we used linear fitting of the measured voltage as a function of applied magnetic field at large magnetic field ranges (near $\pm B_\mathrm{max}$). The exact fitting range depends on the behavior of the measured signal at each gate voltage. A linear fit which takes into account all data points from both up-sweep and down-sweep scans in the respective fitting range (large positive and negative $B_{\parallel}$) is then performed. The fits give two linear curves, $V'_\mathrm{S} = a_{1(2)}B_{\parallel}+b_{1(2)}$, where the subscripts 1 and 2 denote whether the coefficients $a$ and $b$ belong to the large positive or the large negative fitting range, respectively. The difference in the intersections of these two linear fits on the $V_\mathrm{S}$ axis gives $\Delta V_\mathrm{S}$: namely, $\Delta V_\mathrm{S} \equiv b_1 - b_2$. The sign of $\Delta V_\mathrm{S}$ reflects its origin (positive for TSS, negative for Rashba). The error in $b_{1(2)}$ is defined as three times its standard deviation $\sigma_{1(2)}$ in the fit. Hence, the error in $\Delta V_\mathrm{S}$ is specified as $\epsilon_{\Delta V_{\rm S}} = 3(\sigma_{1}+\sigma_{2})$. The process is illustrated in Fig. S2. 

It is also important to note that a clear hysteretic voltage was not always observed in all measurements. At many gate voltages (typically those with near-zero $\Delta V_\mathrm{S}$), the hysteresis was unclear. Nevertheless, applying the aforementioned spin voltage estimation protocol on these data yielded reasonable values of $\Delta V_\mathrm{S}$ (close to 0). Therefore, the same protocol was applied for all $V_{\rm G}$-dependence data.

\section{Ruling out the effect of Nb superconductivity}

\begin{figure}[ht]
    \centering
    \includegraphics[width=0.4\linewidth]{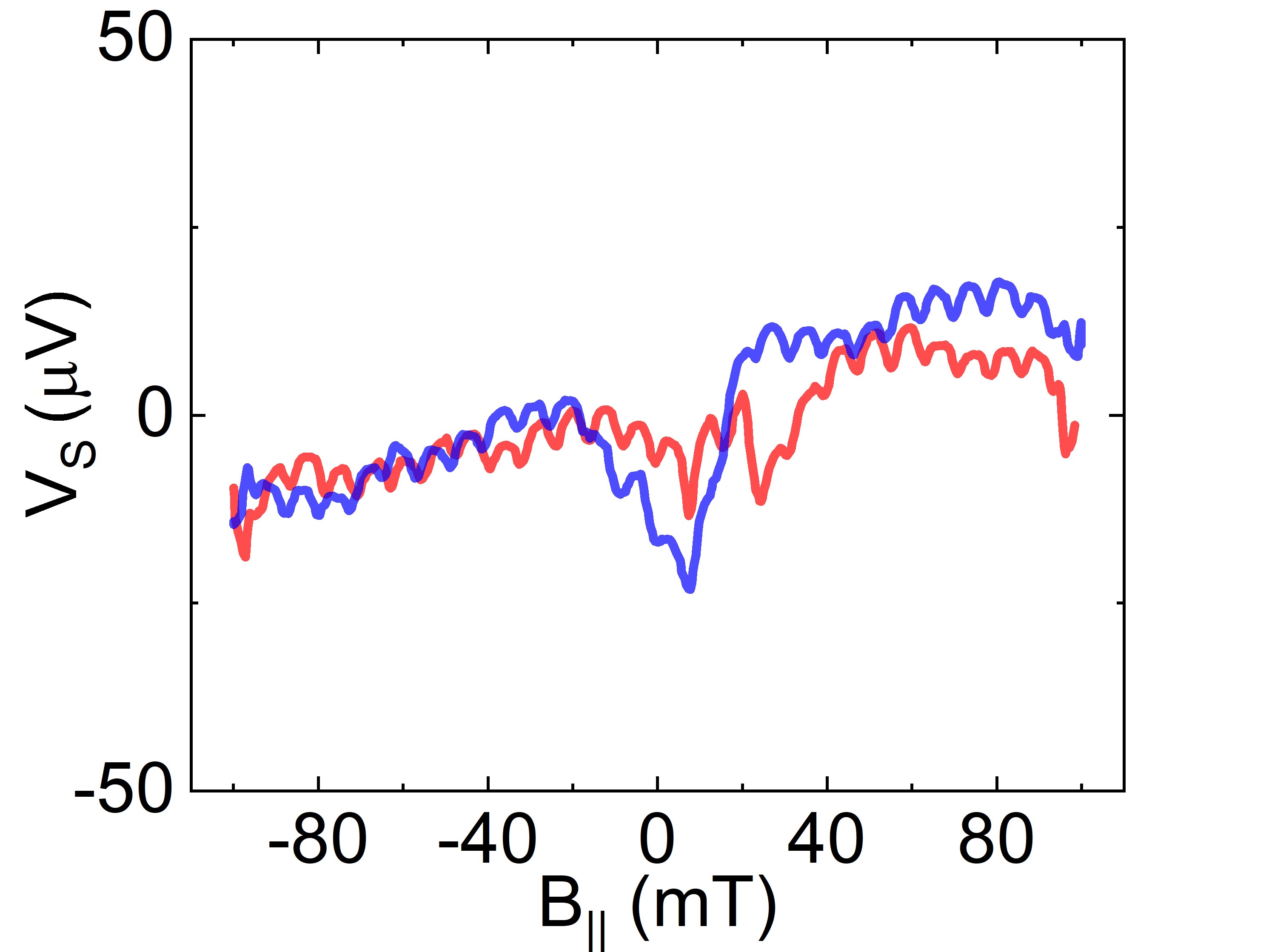}
    \caption[width=\linewidth]{\linespread{1.05}\selectfont{}\label{fig:S3} Voltage hysteresis in device B at 9 K, which is above the $T_c$ of Nb. A finite spin voltage is still discernible in the data.}
\end{figure}

All normal leads of the devices reported in this paper were fabricated using Nb for its good stickiness. However, the spin voltage measurements were performed at the base temperature of the dilution refrigerator (20--40 mK), which is below the superconducting critical temperature $T_\mathrm{c}$ of Nb. To confirm that the superconductivity of Nb does not produce any spurious signal that resembles the spin voltage, we performed the spin voltage measurement at 9 K, which is above the $T_\mathrm{c}$ of our Nb film (which was $\sim$8 K). As shown in Fig. S3, a finite $\Delta V_\mathrm{S}$ is still observed at 9 K, which confirms that the spin voltage is present irrespective of the superconductivity of Nb.

\section{Materials characterization}

The Hall measurement was performed after the spin voltage measurement in each cool-down. Unfortunately, no reliable Hall signal was measured in device A due to bad contacts of the Hall electrodes. The carrier densities estimated from the Hall measurements are shown in Table S1. The carrier density in devices B and C was estimated to be in the high $10^{12}$ cm$^{-2}$ range, which suggests that their chemical potential (at $V_\mathrm{G} = 0$ V) was close to the conduction band edge.

\begin{table*}[h]
    \centering
\begin{tabular}{| c | c | c | c|}
   \hline
\textrm{Device}&
\textrm{A}&
\textrm{B}&
\textrm{C}\\
   \hline
2D carrier density (cm$^{-2})$ & - & 6.72$\times 10^{12}$ & 9.17$\times 10^{12}$\\
   \hline
   TI composition & BiSbTeSe$_{2}$ & Bi$_{0.5}$Sb$_{1.5}$Te$_{1.3}$Se$_{1.7}$ & Bi$_{0.5}$Sb$_{1.5}$Te$_{1.3}$Se$_{1.7}$\\
      \hline
\end{tabular}
\caption{\linespread{1.05}\selectfont{}\label{tab:table1}
Two-dimensional (2D) carrier density estimated from the Hall measurements at $V_{G} =$ 0~V and the composition of the BSTS crystal used for the device. }
\end{table*}

\section{Reproducibility of the spin transistor effect}

\begin{figure*}[h]
    \centering
\includegraphics[width=0.8\linewidth]{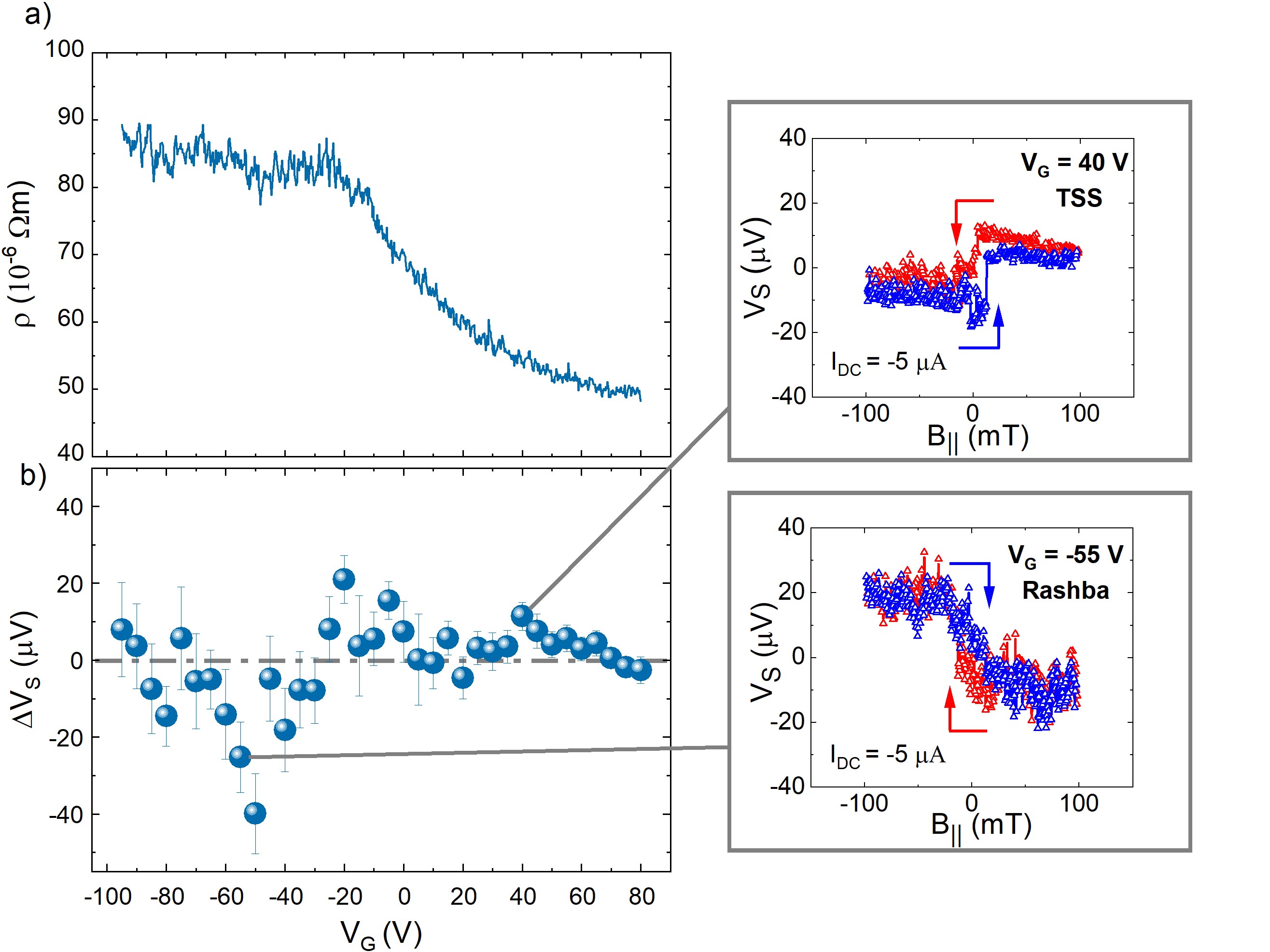}
\caption[width=\linewidth]{\linespread{1.05}\selectfont{}\label{fig:S4} 
a) Resistivity and b) $\Delta V_\mathrm{S}$ as a function of $V_\mathrm{G}$ measured in device C. Insets: opposite spin polarization hysteresis observed between $V_\mathrm{G} = 40$ V and $V_\mathrm{G} = -55$ V. In these measurements of device C, the direction of the DC current was opposite to that in the measurements of device B, and hence the $V_{\rm S}$ vs $B_{\parallel}$ relationship is reversed.}
\end{figure*}

A total of 3 devices are reported in this paper: A, B and C. While device A showed no sign change in the spin voltage $\Delta V_{\rm S}$ upon gating, device B exhibited a sign change and a complicated oscillation in $\Delta V_\mathrm{S}$ as a function of $V_\mathrm{G}$. The spin voltage reversal was reproduced in device C: Not only the opposite hysteresis was observed between $V_\mathrm{G} = 40$ V (TSS origin) and $V_\mathrm{G} = -55$ V (Rashba origin), but also the oscillation between plus and minus $\Delta V_\mathrm{S}$ as a function of $V_\mathrm{G}$ was reproduced (see Fig. S4).  This reproducibility gives confidence in the intrinsic nature of the spin transistor effect in our TI-based devices.

\section{Silicon nitride tunnel barrier}

The impedance mismatch between a semiconductor (which is TI in our case) and a ferromagnetic (FM) electrode can strongly hinder the detection of the spin polarization using a FM potentiometer \cite{mismatch}. Therefore, a tunnel barrier between the spin-active material (TI) and the FM is required to compensate for this mismatch \cite{rashba2000}. Experimentally, reproducible fabrication of a tunnel barrier with suitable resistance is the main challenge for this type of devices. 
In this context, Al$_{2}$O$_{3}$ deposited by ALD is the material conventionally used for this tunnel barrier due to its high dielectric constant and well-controlled thickness. However, the resistivity of Al$_{2}$O$_{3}$ deposited by ALD was not reproducible enough in our experiment. (We observed a large variation in the tunnel barrier resistance ). One possible reason for this irreproducibility is the native oxide on the TI surface that is uncontrolled. 

We therefore employed SiN$_{x}$ tunnel barrier fabricated with a hot-wire chemical vapor deposition (HW-CVD). This technique was used because of the low substrate temperature during the deposition \cite{Yang2014}, in contrast to the high substrate temperature typically associated with the CVD technique. This low substrate temperature makes HW-CVD compatible with TI devices, since experiencing a temperature above 120$^{\circ}$C will shift the chemical potential in BSTS \cite{Breunig2022}. 
The SiN$_{x}$ barriers deposited by HW-CVD showed more reproducible tunnel-barrier resistances compared to the ALD-prepared Al$_{2}$O$_{3}$ tunnel barrier. 

\begin{flushleft}
{\bf References}
\end{flushleft}

\providecommand{\noopsort}[1]{}\providecommand{\singleletter}[1]{#1}%